\documentclass{Interspeech}
\usepackage{placeins}
\newcommand{\e}{\mathrm{e}}
\usepackage{comment}

\interspeechcameraready

\usepackage{comment}
\begin{document}

\title{Articulatory modeling of the S-shaped F2 trajectories observed in Öhman's spectrographic analysis of VCV syllables}

\author[affiliation={1}]{Frédéric}{Berthommier}


\affiliation{Univ. Grenoble Alpes, CNRS}{Grenoble INP, GIPSA-Lab}{France}

\email{frederic.berthommier@gipsa-lab.grenoble-inp.fr}

\maketitle








\begin{abstract}
The synthesis of Öhman's VCV sequences with intervocalic plosive consonants was first achieved 30 years ago using the DRM model. However, this approach remains primarily acoustic and lacks articulatory constraints. In this study, the same 75 VCVs are analyzed, but generated with the Maeda model, using  trajectory planning that differentiates vowel-to-vowel transitions from consonantal influences. Synthetic data exhibit similar characteristics to Öhman’s sequences, including the presence of S-shaped F2 trajectories. Furthermore, locus equations (LEs) for F2 and F3 are computed from synthetic CV data to investigate their underlying determinism, leading to a reassessment of conventional interpretations. The findings indicate that, although articulatory planning is structured separately for vowel and consonant groups, S-shaped F2 trajectories emerge from a composite mechanism governed by the coordinated synergy of all articulators.
 \end{abstract}

\section{Introduction}

The S-shaped transition of the second formant (F2) in a V\textsubscript{1}CV\textsubscript{2} sequence, where V\textsubscript{1} and V\textsubscript{2} have opposite characteristics (front/back), is a key element for understanding the phenomenon of coarticulation. To date, this transition has not been systematically modeled using an articulatory framework. Addressing this issue—which extends beyond the well-known many-to-one articulatory-to-acoustic relationship—requires additional constraints. Some of these constraints may be biomechanical, arising from articulation costs, while others could be imposed by an articulatory model with well-characterized articulatory-to-acoustic properties, such as the Maeda model \cite{Maeda1990}. One challenge is that most studies rely on articulographic trajectories, which, for tongue movements, do not necessarily align with the goals or articulatory parameters of Maeda’s model. The articulatory basis of S-shaped F2 transitions remains poorly understood, despite their crucial role in revealing coarticulation phenomena, and we aim to revisit this issue. Initially, Öhman \cite{Ohman1967} concluded with his modeling approach that 
\begin{quote}
    "Vowel and consonant gestures are largely independent at the level of neural instructions." 
\end{quote}
This hypothesis was developed later by \cite{Perkell1969} with an anatomical point of view. He subdivided the vocal tract musculature into two tiers: one for vowels and the other for consonants. Mathematical models of area function variations \cite{Story2005}, successfully applied this idea with two overlapping functions. The vowel tier corresponds to slow, global shape variations of the vocal tract between V\textsubscript{1} and V\textsubscript{2}. Conversely, the consonant tier involves rapid, localized variations of the vocal tract, aiming to establish a constriction site. This mechanism produces convincing perturbations of the formant trajectories and fulfills the concept of an independent control of vowels and consonants, but unfortunately does not provide a representation in terms of articulatory parameters. The VocalTractLab model provides an implementation \cite{Birkholz2013} inspired by \cite{Ohman1967}, built on variations of a human-like vocal tract. For the synthesis of CV trajectories, this model weights three basic shapes for each consonant (/C\textipa{a}/, /C\textipa{i}/, and /C\textipa{u}/) according to the vowel’s position in the vowel triangle. However, it remains primarily focused on practical speech synthesis rather than on verifying knowledge about coarticulation. Öhman's S-shaped transitions correspond to gradual yet nonlinear changes in formant trajectories during vowel-consonant coarticulation. However, by using segmental control over eight regions of a closed-open tube, the DRM \cite{Mrayati1988} naturally generates these sigmoid trajectories without the need to adjust articulatory parameters \cite{CARRE1995, Chennoukh1997, CARRE2017}. After a long period of dominance by the AP/TD framework \cite{Browman1992, Saltzman1989} in this field, \cite{Xu2025} is reintroducing temporal constraints on articulators to regulate coarticulation, particularly in terms of the synchronization required for syllable production and the modeling of S-shaped transitions. This approach notably discards the use of coupled oscillators from the AP/TD framework for establishing articulatory phasing, favoring simpler mechanisms instead. According to \cite{Xu2025}, Öhman's S-shaped trajectories emerge from the synchronization between the vowel and consonant tiers. Since our mathematical modeling of syllable planning enables the reproduction of all results obtained with the DRM using the Maeda model, including those related to Öhman's spectrographic measurements and locus equations, this opens the opportunity to reinterpret these two key observations in the context of a true articulatory model. Parameter coordination is achieved through a purely mathematical framework that integrates the kinematic formulation of the Tau model \cite{Lee2009}, which has been successfully applied to articulographic data \cite{Elie2023}. 
\section{Syllable synthesis with the Maeda model}
\begin{figure*}[t]
\centering
\includegraphics[scale=0.50]{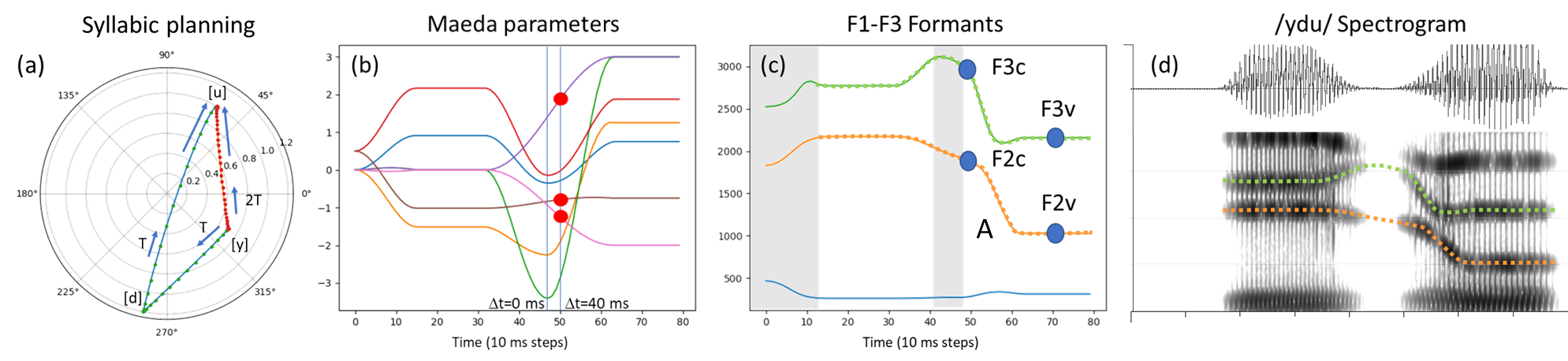}
\caption{Syllable planning and synthesis of the utterance /ydu/.}
\label{fig:Principle}
\end{figure*}

\subsection{Planning of trajectories}

The coarticulation between vowels and consonants in VCV sequences was modeled by \cite{Ohman1967} using an equation that describes the variations of the vocal tract shape along the glottis-to-lips axis \(x\). A reformulation of this equation reveals that it represents a blending process between a time-dependent vocalic component \(v(x; t)\) and a static consonantal component \(c(x)\). The kinetic term \(k(t)\), which varies from \(0\) to \(1\) and then from \(1\) to \(0\), evolves relatively quickly, while the coarticulation function \(w_c(x)\) determines the localization of the constriction along the \(x\)-axis:
\begin{align}
\begin{cases}
s(x; t) = v(x; t) + k(t) \, (c(x) - v(x; t)) \, w_c(x) \\
\phantom{s(x; t)} = (1 - k(t) \, w_c(x)) \, v(x; t) + k(t) \, w_c(x) \, c(x)  \\
v(x; t) = (1 - k'(t)) \, v_1(x) + k'(t) \, v_2(x)
\end{cases}
\label{eq1}
\end{align}
The slow variation process is governed by its own monotonic kinematic term, \(k'(t)\), which transitions smoothly from \(0\) to \(1\), defining a trajectory between the two static vowels, \(v_1(x)\) and \(v_2(x)\). Note that the two equations together describe trajectories that transition from \(v_1(x)\) toward the two targets, \(c(x)\) and \(v_2(x)\). The principle of superimposing slow and fast variations for vowels and consonants has been successfully applied within the DRM framework \cite{CARRE1995, Chennoukh1997} to replicate the formant modulations observed by \cite{Ohman1966}. This approach has also been employed to introduce specific constrictions superimposed onto the area function of slowly varying vowels \cite{Story2005, Story2024}. However, extending this principle to articulatory models presents greater challenges, as control is not directly applied along the \(x\)-axis but rather within a more abstract articulatory space.

This principle was later subsumed under Fowler's coproduction framework \cite{Fowler1980}, which serves as a foundation for articulatory phonology. This framework assumes that gestures are relatively independent yet temporally constrained. The type of control proposed by Task Dynamics was recently implemented by \cite{Alexander2019}, but, to date, there is no explicit modeling of S-shaped transitions as in \cite{CARRE1995, Chennoukh1997}. Nevertheless, these implementations diverge significantly from the original formulation. We propose to revisit and reinstate the application of this principle in articulatory modeling, adhering more closely to its initial conceptualization.

In the Maeda model, Öhman's shape terms, \(c(x)\), \(v(x; t)\), and \(s(x; t)\), are inherently factorized through Principal Component Analysis (PCA) to extract articulatory parameters. By design, these parameters are independently controlled and can be assigned to either the slow vocalic or fast consonantal variation processes. Consequently, the coarticulation term \(w_c(x)\) becomes unnecessary. 

First, the control of parameter variation is conducted in the polar plane, enabling temporal modulation of values rather than merely activating or deactivating articulators, as in the gestural score of articulatory phonology. This guarantees continuous articulator activity, leveraging agonist-antagonist relationships—fundamental properties of the Maeda model \cite{Maeda1994}. This approach yields a set of three equations, with kinetic terms defined along three oriented arcs (Figure~\ref{fig:Principle}a): one connecting each vowel to the consonant and another directly linking the two vowels. These arcs have respective durations of \( T \) and \( 2T \): \( (V_1, C, T) \), \( (C, V_2, T) \), and \( (V_1, V_2, 2T) \). As a consequence, the \( V_1 V_2 \) transition is twice as slow as the \( V_1 C \) and \( C V_2 \) transitions, ensuring synchronization between the two pathways. This planning applies equally to both VCV and CV structures, as the latter begins with an anticipatory reduced vowel \( \textipa{V}_{\text{red}}CV \).
\FloatBarrier
\begin{figure*}[htbp] 
\centering
\includegraphics [scale=0.50] {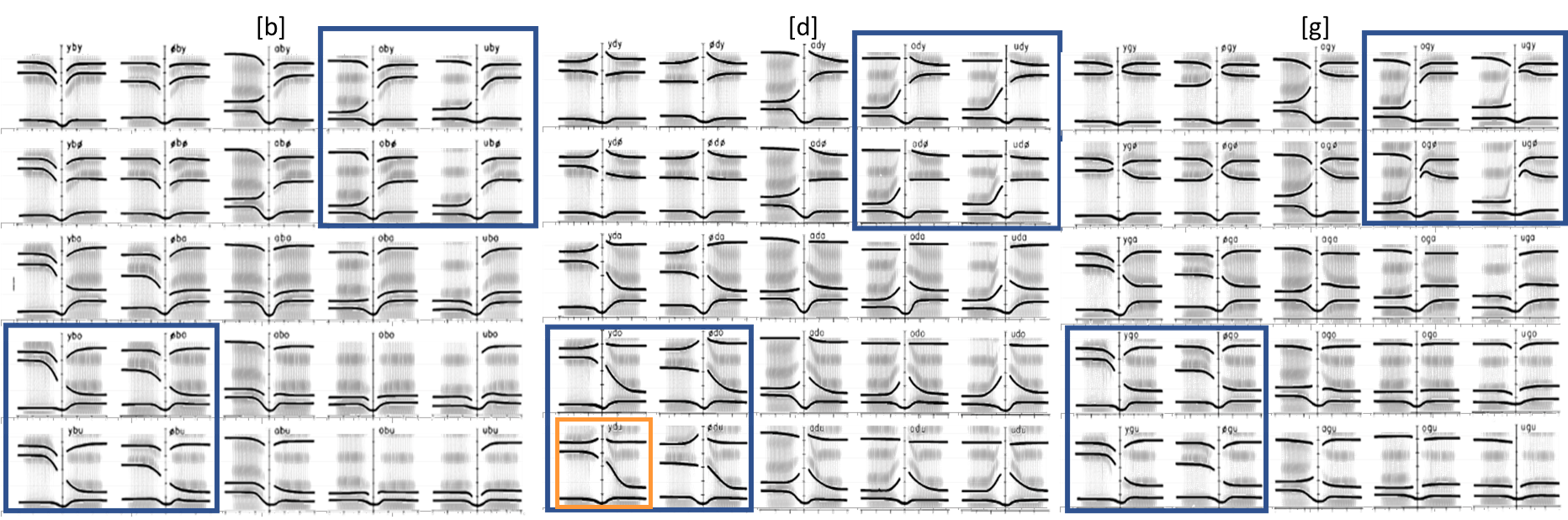} 
\caption{Successful superimposition of Öhman 1966 \cite{Ohman1966} and spectrograms of synthetic VCVs.}
  \label{fig:Ohman}
\end{figure*}

In the syllable planning polar plane, nodes are phonemes and arcs are joining pairs of nodes to form a syllable graph. Trajectories for vowels and consonants are planned in the complex plane by forming arcs between 2 points $(\rho_1,\theta_1)$ and $(\rho_2,\theta_2)$:
\begin{align}
  z(t) = \rho(t) \, \rho_1 \, \e^{i \theta_1} + (1-\rho(t)) \, \rho_2 \, \e^{i (\theta_2 + \frac{\nu}{K} \theta(t))} 
 \label{eq2}
\end{align}
where $t$ varies between $1$ and $nT$, $\rho(t)$ determines the velocity profile and $\nu=\pm1$, $\theta(t)$ and $K$ are supplementary shape parameters. When $K$ is large ($K=30$ for vowel arcs and $K=10$ for consonant arcs), the trajectories are quasi-straight lines (see Figure~\ref{fig:Principle}a). Otherwise arcs are curved to generate useful variability as in \cite{Iskarous2010b}. The positions \((\rho_i, \theta_i)\) of vowels and consonants in the complex plane serve as an abstract representation of their place of articulation and degree of opening. This model-flexible encoding provides an alternative to specifying physical targets along the vocal tract, as done in the AP/TD framework. This key feature enables the application of the same graph structures across different vocal tract representations. To produce the arcs of Figure 1a for the syllable /ydu/, for each phonetic target, we have [\textipa{y}]: \((0.7, \frac{11\pi}{6})\), [\textipa{d}]: \((1.2, \frac{23\pi}{16})\), and [\textipa{u}]: \((1, \frac{\pi}{3})\). The parameters specific to each arc are assigned in order to compute their variations. Maeda’s model has seven parameters \{Jaw, Body, Dorsum, Tip, LipP, LipH, Hy\}, denoted from 1 to 7, and the consonantal parameters are for [\textipa{b}]: \{1,2,6\}, [\textipa{d}]: \{1,2,3,4\}, and [\textipa{g}]: \{1,2,3,4\}. The remaining parameters are assigned to the vocalic arc, with vowels incorporating all ones. It is observed that the jaw (=1) is always recruited to complete the consonantal occlusion and that the Body parameter (=2) is involved in [\textipa{b}] in order to produce the trough effect \cite{Lindblom2002}. All tongue parameters are involved in the production of [\textipa{d}] and [\textipa{g}], and this was chosen experimentally, which may contradict Öhman's simulation results \cite{Ohman1967}.

\subsection{Incorporation of the Tau model}
This design is compatible with the Tau model reported and tested in \cite{Elie2023}, which enables the attainment of consonant and vowel targets within a specified duration, either \( T \) or \( 2T \). The core equation of this model describes a continuous decay of \( X(t) \) from an initial value \( X_0 \) to zero as \( t \) progresses from \( 1 \) to \( D \): 
\begin{align}  
X(t) = X_0 \left( 1 - \frac{t^2}{D^2} \right)^{\frac{1}{\kappa}}  
\label{eq3}  
\end{align}  
A closely related harmonic form (additional details are provided in the supplementary materials \cite{BerthommierSM} for verification of the following claims) describes the harmonic decay over a quarter period, where \(t \in [1, D]\):  
\begin{align}  
X(t) = X_0 \cos\left(\frac{\pi t}{2D}\right)^{\frac{1}{\kappa}}  
\label{eq4}  
\end{align}  
In the trajectory equations, this latter form is applied to \(\rho(t)\) with \(X_0 = 1\) in Eq.~\ref{eq2} but for the \( V_1 C \) transition only. For the other two, \(\rho(t) = 1 - X(D-t)\) is used instead, which is a variant that preserves the principle of gap closure. In Eq.~\ref{eq4}, \(\kappa = 0.5\) is chosen to closely approximate the behavior of Eq.~\ref{eq3} with \(\kappa = 0.4\) (the median value identified by \cite{Elie2023} using articulographic data). Furthermore, the new forms (Eq.~\ref{eq4} and variant) are equivalent at this \(\kappa\) value, evolve smoothly from their initial values, and converge precisely to the target values at \( t = D \). In syllable planning, slow vowel-to-vowel transitions follow \(D = 2T\), while faster consonantal transitions adhere to \(D = T\). In the following simulations \( T = 16 \) with \(\SI{10}{\milli\second}\) steps.

\section{Modeling the S-shaped transition}

The 75 syllables V\textsubscript{1}CV\textsubscript{2}, composed of the vowels V\textsubscript{1} and V\textsubscript{2}, were selected from the set \textipa{/y, \o, a, o, u/}. Each vowel is assigned a position on the unit circle: [\textipa{y}]: \((0.7,\frac{11\pi}{6})\), [\textipa{\o}]: \((0.3,\frac{11\pi}{6})\), [\textipa{a}]: \((0.8,\pi)\), [\textipa{o}]: \((0.8,\frac{\pi}{2})\), and [\textipa{u}]: \((1,\frac{\pi}{3})\). Similarly, the consonants are positioned independently of the vowels, slightly outside the unit circle to generate a constriction at an appropriate place of articulation: [\textipa{b}]: \((1.2,\frac{\pi}{3})\) and [\textipa{d}]: \((1.2,\frac{23\pi}{16})\). The consonant [\textipa{g}\textsubscript{p}] is associated with front vowels \textipa{/y, \o/} and is placed at \((1.1,\frac{23\pi}{12})\), whereas [\textipa{g}\textsubscript{v}] is associated with back vowels \textipa{/a, o, u/} and is positioned at \((1.2,\frac{\pi}{3})\). Notably, the placement of [\textipa{g}\textsubscript{v}] coincides with that of [\textipa{b}]; however, their consonantal parameters differ (as detailed above).  
For the transition portion of each utterance, the Maeda parameters  (Figure~\ref{fig:Principle}b) are derived from the 3 aggregated trajectories using the values \(\bm{\Omega}\) and \(\bm{\Psi}\) given in Suppl. S1 (see \cite{BerthommierSM}): 
\begin{align}
    \bm{P}(t) - \bm{\Omega} = \text{Re} \left[ S_v \odot \bm{\Psi} \bar{z}_v(t) + S_c \odot \bm{\Psi} \bar{z}_c(t) \right]
    \label{eq5}
\end{align}  
where \( S_v \) and \( S_c \) are mutually exclusive selection vectors, composed of zeros and ones, specifying the activation of vocalic and consonantal articulators. These vectors are applied element-wise (Hadamard product) to the complex vector \( \bm{\Psi} \), ensuring the appropriate modulation of the articulatory parameters.
The F1–F3 formants (Figure~\ref{fig:Principle}c) are computed using a classical transmission line method \cite{BadinFant1984}. The source model remains simplistic, with minimal post-processing applied to the synthesized output. Specifically, the signal is multiplied by an amplitude envelope that depends on the VCV syllable structure, introducing a silent period around the consonant onset center in gray (Figure~\ref{fig:Principle}c). The S-shaped characteristic of the F2 transition is superimposed on the spectrogram Figure~\ref{fig:Principle}d, demonstrating that it is not strictly equivalent to the sigmoid transition of the vocalic parameters, indicated by red dots in Figure~\ref{fig:Principle}b. This confirms the presence of an interaction between the two parameter groups we will describe later.

\begin{figure}[b!]
  \centering
  \includegraphics[width=\linewidth]{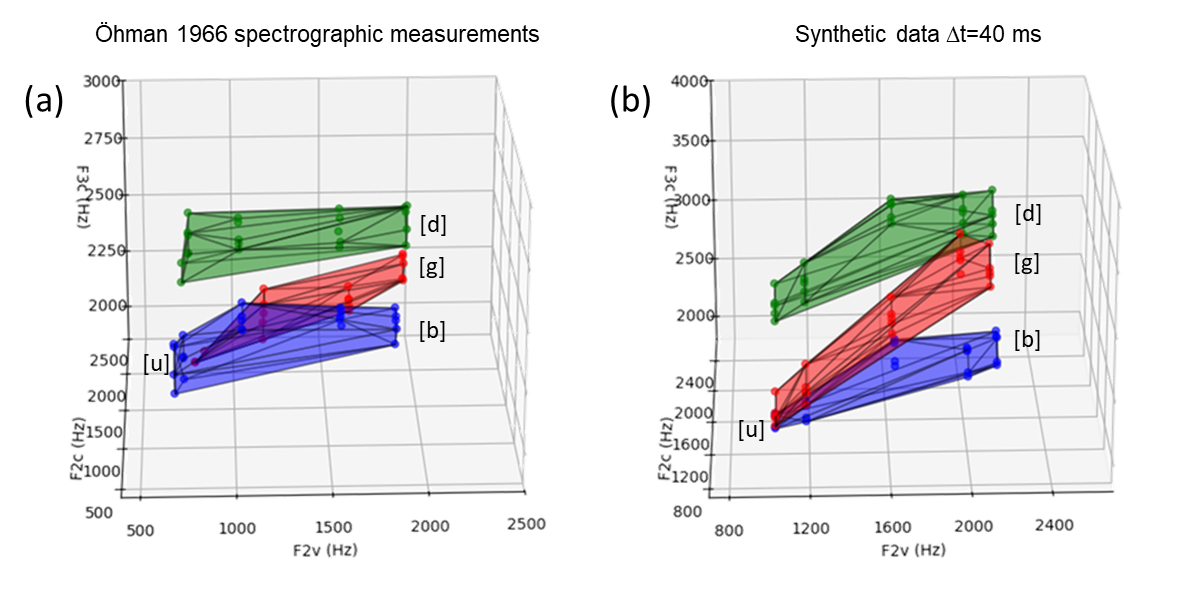}
    \caption{3D representation as in Lindblom et al. \cite{Lindblom2010}.}
  \label{fig:Lindblom}
\end{figure}
In Figure~\ref{fig:Ohman}, aligned and superimposed representation of synthetic and observed data: original Öhman’s figures \cite[Fig. 7, 8 and 9]{Ohman1966} are manually aligned and superimposed with 75 synthetic spectrograms corresponding to the three consonants [\textipa{b}], [\textipa{d}], and [\textipa{g}]. Blue squares highlight the front-to-back and back-to-front vowel-consonant-vowel sequences, which exhibit typical S-shaped F2 trajectories. Otherwise, consonants appear as perturbations in vowel trajectories.

For a concise comparison, as suggested by \cite{Lindblom2010}, we utilized Öhman's data \cite{Ohman1966} from Table II and Table IV, recombining them to form clusters for each consonant in the 3D Figure~\ref{fig:Lindblom}a with axis \(x\)=F2v, \(y\)=F2c and \(z\)=F3c. In Figure~\ref{fig:Lindblom}b, we extracted from the F2-F3 formant modulation of the synthetic data these 3 points indicated in blue Figure~\ref{fig:Principle}c with F2c and F3c at a \(\Delta t\) of \SI{40}{\milli\second} from the consonant onset center at \(\Delta t\)=\SI{0}{\milli\second}. While the amplitude of F3 variations is greater in the synthetic data because of the geometry of the Maeda model, the overall structure remains consistent. The Figure~\ref{fig:Lindblom} is oriented to display the F2v-F2c relationship in perspective which is similar to the locus equations \cite{Sussman1991} established for the CV transition. In both cases, the [\textipa{b}] and [\textipa{g}] clusters overlap for the vowel [\textipa{u}], preventing full separability with these features only, in contradiction with \cite{Lindblom2010}. Alternatively, in the absence of a burst for /V\textipa{gu}/, the indentification of [\textipa{g}] might be allowed by the /V\textipa{g}/ transition.

\begin{figure}[b!] 
\centering
  \includegraphics[width=\linewidth]{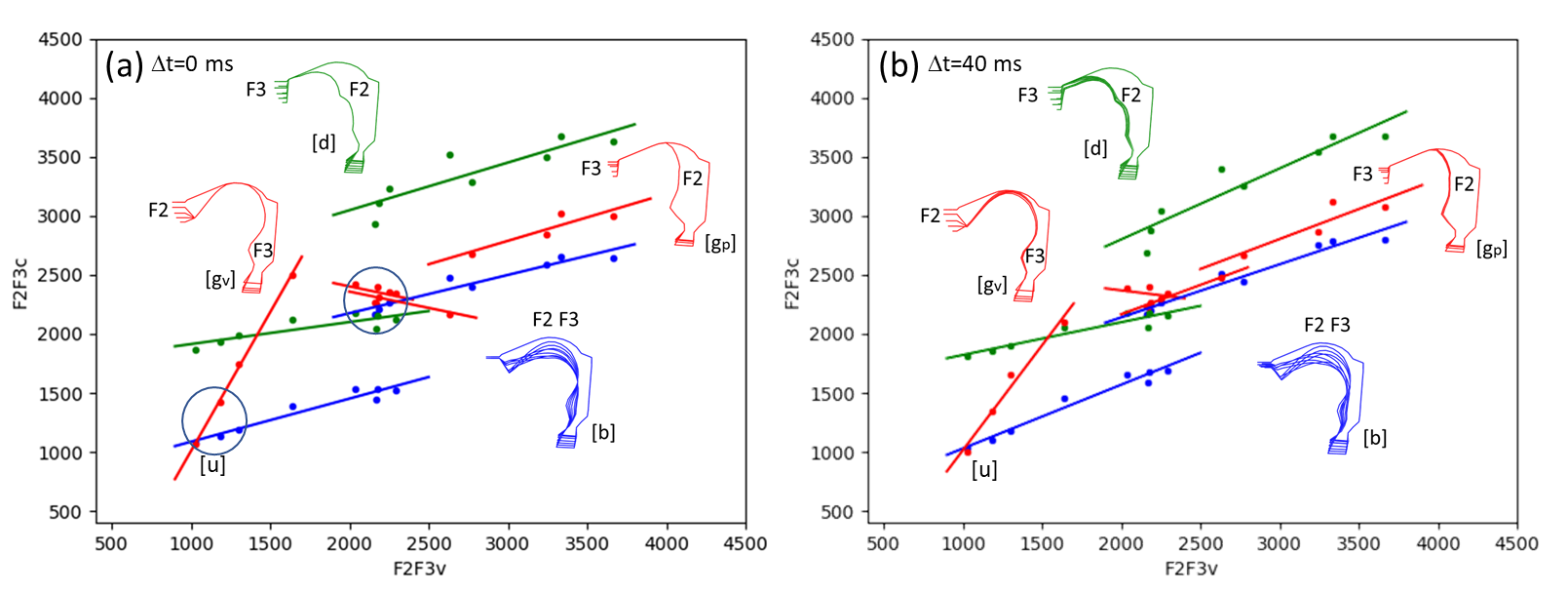}
  \caption{Comparision of F2-F3 locus equations at \(\Delta t\) of \SI{0}{\milli\second} and \SI{40}{\milli\second} as in Bailly \cite{Bailly1995} for showing formant affiliations.}
  \label{fig:Delay}
\end{figure}

\section{Where do S-shaped F2 transitions originate?}
To provide a physical interpretation of locus equations Bailly \cite{Bailly1995} focused on tracking the resonances of the vocal tract, based on the premise that formants are associated with the front and back cavities. When a constriction occurs in the vocal tract, the resonant cavities are decoupled, allowing formants to be affiliated to specific cavities. This reasoning contrasts with perturbation theory, which models deformations of a neutral tube, as in \cite{Mrayati1988, Story2005}. Supplementary Figure S2a (in \cite{BerthommierSM}) from \cite{Bailly1995} allows this previously overlooked interpretation of the locus equations for [\textipa{g}] to be revisited, distinguishing between palatal [\textipa{g}\textsubscript{p}], associated with front vowels, and velar [\textipa{g}\textsubscript{v}], associated with back vowels. The experimentally observed relationships between F2v-F2c and F3v-F3c \cite{Klatt1987} reveal that F2 of [\textipa{g}\textsubscript{v}] aligns with F3 of [\textipa{g}\textsubscript{p}]. This occurs because F2 is affiliated with the oral cavity for back vowels, while F3 is similarly associated for front vowels. The  linearity of the F2v-F2c locus equation emerges not only from tongue position but also from the degree of lip opening. 

The modeling of CVs with the eight vowels /\textipa{u, o, O, a, E, e, i, y}/, following the same procedure as for VCVs but with \( \textipa{V}_{\text{red}}CV \), is obtained by taking the formant values at a \(\Delta t\) of \SI{0}{\milli\second} and \SI{40}{\milli\second}. The position of \( \textipa{V}_{\text{red}} \) in the unit circle is $(0.5 \rho_V,\theta_V)$. Continuing the analysis of [\textipa{g}], the state of the vocal tract and formant affiliations are illustrated in Figure S2b (see \cite{BerthommierSM}) and Figure~\ref{fig:Delay}a-b, highlighting that LEs are also governed by vowel parameters: \(\text{LipP}\) and \(\text{LipH}\) for the front cavity, and \(\text{Hy}\) for the back cavity. 
Globally, at the consonant onset center \(\Delta t = \SI{0}{\milli\second}\) in Figure~\ref{fig:Delay}a, the slopes of the F2v-F2c LEs are notably steep for [\textipa{b}] and [\textipa{g}\textsubscript{v}], challenging the hypothesis of an invariant consonant locus \cite{Delattre1955}, which was later reinstated by \cite{Lindblom2012}. However, this condition is only met when the slope is flat, as observed for [\textipa{d}], where F2 remains affiliated with the back cavity and is influenced solely by \(\text{Hy}\). At \(\Delta t = \SI{40}{\milli\second}\) in Figure~\ref{fig:Delay}b, the tongue moves toward the vowel target, aligning with the linear correlation between \(\text{TBx}\) at \(V\) and \(\text{TBx}\) at \(C\), as observed by \cite{Iskarous2010}. Due to the kinematics of the Tau model, the speed of movement at consonant release is relatively low (see Figure~\ref{fig:Principle}b), suggesting that this factor contributes moderately to the slope of the LEs.

The observed non-linearity of F2 arises from the coordinated interaction of all parameters and is not merely the result of a sigmoidal variation in vowel parameters that produces a diphthongal transition, as suggested by \cite{Perkell1969}. First, as illustrated in Figure~\ref{fig:Principle}b-c, the effective V-to-V transition begins with the consonant release, with the tongue moving towards [\textipa{u}]. The articulation of the three phonemes and their coarticulation follow a sequential scheme: \(V_1 \rightarrow\) Constriction \( \rightarrow \) Transition to \(V_2 \rightarrow V_2\). To analyze the six /\textipa{u}C\textipa{y}/ and /\textipa{y}C\textipa{u}/ combinations, we introduce an additional Figure S3 (in \cite{BerthommierSM}) that relates the timing of the transition to its structural planning in the polar plane. To ensure consistently straight trajectories and avoid introducing additional nonlinearity, we set \( K = 1000 \) in Eq.~\ref{eq2}.

When the main part of the F2 transition occurs after \(\Delta t = \SI{0}{\milli\second}\), it is classified as case A as in Figure~\ref{fig:Principle}c; conversely, when it occurs before, it is classified as case B. If the arrows along the first half of the vowel pathway (with duration \(T\)) and along the \(V_1 C\) arc converge towards \(V_2\), the transition occurs before the constriction. This pattern is symmetric for the case A transitions as in Figure~\ref{fig:Principle}a. Thus, the converging directions of the two pathways govern the synergy, making the phenomenon independent of formant affiliation and instead driven by relative positions within the polar plane. Considering that these positions themselves are linked to places of constriction of consonant and vowels provides an intuitive explanation of the underlying mechanism. In a \(V_1 C V_2\) syllable with two distant vowels, when \(V_1\) is closer to \(C\), the transition occurs after the consonant constriction; conversely, when \(C\) is closer to \(V_2\), the F2 transition occurs before the consonant constriction. As a corollary, when the two vowels are close or identical, the F2 trajectory appears to be disrupted by the consonant closure.

The S-shaped F2 trajectory emerges from the relationship between five key points: F2v\textsubscript{2}, F2c\textsubscript{2} at \(\Delta t = \SI{40}{\milli\second}\), F2 at the onset center, F2c\textsubscript{1}  at \(\Delta t = -\SI{40}{\milli\second}\) along with F2v\textsubscript{1}. However, can interpretable insight be derived from the locus equations ? A close analysis of Figure~\ref{fig:Delay}a reveals that the two circles lie near the diagonal, indicating that for a CV sequence, F2 remains relatively stable for [\textipa{d}] and [\textipa{g}\textsubscript{p}] when V is a front vowel, meaning that the transition occurs before in a \(V_1 C V_2\) when \(V_1\) is a back vowel. Symetrically, for [\textipa{b}] and [\textipa{g}\textsubscript{v}] with V as a back vowel, the F2 transition of \(V_1 C V_2\) is also a case B when \(V_1\) is a front vowel. For [\textipa{b}] and [\textipa{d}] the cases A correspond to the opposite V vowels (2/6 in Figure S3 \cite{BerthommierSM}), which are far from the diagonal and exhibit a significant difference between F2c and F2v, due to the F2 transition.

\section{Conclusion}
From simulations of syllable structure using an articulatory model, we derive plausible deterministic principles that account for key historical observations of the coarticulation phenomenon. All simulations and figures are fully reproducible using the open-source Python software provided with this work \cite{berthommier2025}. This approach extends articulatory modeling beyond earlier simulations using tube models \cite{CARRE1995,Chennoukh1997,CARRE2017} and area functions \cite{Story2005}, offering deeper insight into articulatory mechanisms and syllable planning. While the model is a simplified representation—currently limited to plosives—it offers a clear advantage in interpretability regarding speech production principles. However, its realism is limited, as it is not data-driven and lacks observable control mechanisms. A promising direction for future work is to explore how syllabic planning might be inferred from articulatory data. An open question is how this structural knowledge could be integrated with learned syllabic representations, as in the Sylber framework \cite{cho2025}.

\section{Acknowledgments}
I gratefully thank \textbf{Gérard Bailly} for his insightful explanations regarding the physical interpretation of the locus equations based on the resonances of the vocal tract. His input was essential for connecting the acoustic patterns with the underlying articulatory mechanisms. I also thank \textbf{Louis-Jean Boë} for personally sharing the MATLAB code of VLAM, which served as the basis for a direct Python translation used in this work.

\bibliographystyle{IEEEtran}
\bibliography{mybib}

\end{document}